\documentclass[twoside,12pt]{article}
\usepackage{epsfig}

\newcommand{\be}{\begin{equation}}
\newcommand{\ee}{\end{equation}}
\newcommand{\bea}{\begin{eqnarray}}
\newcommand{\eea}{\end{eqnarray}}

\begin{document}
\title{ \vspace{1cm} Flavor Symmetries, Neutrino Masses and Neutrino
Mixing\footnote{Invited talk given at the 4th International Conference
on Flavor Physics, Peking (JCFP 2007), September 2007}}
\author{H.\ Fritzsch, Universit\"at M\"unchen, Department f\"ur Physik,
Munich, Germany}
\maketitle
\begin{abstract}
{We discuss the neutrino mixing, using the texture 0 mass matrices, which
work very well for the quarks. The solar mixing angle is directly linked
to the mass ratio of the first two neutrinos. The neutrino masses are
hierarchical, but the mass ratios turn out to be much smaller than for
quarks. The atmospheric mixing angle is 38$^{\circ}$. The CP violation
for leptons should be much smaller than for quarks.}
\end{abstract}
The central problem in flavor physics is a deeper understanding of the
quark and lepton masses. Thus far we only understand the hadron masses,
e. g. the mass of the nucleon. These masses are generated in
QCD\cite{Fritzsch} by the quark--gluon interaction. The confinement
of the colored quarks and gluons leads to the appearance of a mass for the
nucleons and the
other hadrons. This mass is the confined field energy of the gluons and
quarks inside the hadrons. The scale parameter $\Lambda_c$ of QCD,
measured to be about 220 MeV, determines the hadronic masses. But the masses
of hadrons, especially the heavy ones like the $J/\psi$--meson, depend
also on the quark masses, and they remain mysterious.

The Standard Model
has 28 fundamental parameters, 22 of these parameters are directly related
to the fermion masses: 6 quark masses, 3 charged lepton masses,
3 neutrino masses, 4 flavor mixing parameters for the quarks, and 6 for
the leptons.

In the Standard Model the masses of the $W/Z$--bosons are due to the
''Higgs``--mechanism, but we still do not know, whether the
``Higgs``--model\cite{Higgs} is true. Soon we shall find out, when the
new LHC--accelerator at CERN starts producing experimental data.

It remains open, whether there exist relations between the
fermion mass parameters. Many years ago I proposed such relations between
the quark masses and the flavor mixing angles\cite{Fri}. Using the
parametrization, given in ref. (4), these relations are:
\begin{eqnarray}
\Theta_u & = & \sqrt{m_u/m_c} \nonumber \\
\Theta_d & = & \sqrt{m_d / m_s} \, .
\end{eqnarray}
The Cabibbo angle is approximately given by
\begin{equation}
\Theta_c \cong \mid \sqrt{\frac{m_d}{m_s}} - e^{i \phi} \sqrt{\frac{m_u}{m_c}}
\mid
\end{equation}

Taking into account the recent experimental data, these relations work
very well. Similar relations might also exist for the leptons, as discussed
below.

I shall concentrate on the neutrino mixing. About 10 aears ago Xing and
I\cite{Xifri} discussed the possibility that the mixing angles for the leptons
are large, even maximal. The recent data support this hypothesis. But it is
still unclear, what type of masses the neutrinos have.  Are these masses
like the masses of the charged leptons, i. e. Fermi--Dirac--masses? Or are
they Majorana masses? In any case these masses are very small, probably
less than 1 eV.  In the Standard Model with Fermi--Dirac neutrino masses
this is not understood. If the neutrino masses are Majorana masses, one
can introduce these masses, using the see--saw mechanism\cite{Min}. If the
Standard Model is embedded in the Grand Unified Theory,
based on SO(10)\cite{Frimin}, the small Majorana neutrino masses
reflect the heavy masses
of the righthanded neutral leptons, which are part of the
16--dimensional fermion representation of $SO(10)$.

The relations (1) follow from an underlying texture zero mass matrix:
\begin{equation}
M = \left( \begin{array}{lll}
O & A & O \\
A^* & D & B \\
O & B^* & C
\end{array} \right)
\end{equation}

We describe the neutrino mixing by the following flavor mixing matrix:
\begin{equation}
V = U \cdot P
\end{equation}
\begin{equation}
P = \left( \begin{array}{ccc}
e^{i \rho} & 0 & 0 \\
0 & e^{i \sigma} & 0 \\
0 & = & 1
\end{array} \right)
\end{equation}
\begin{equation}
U = \left( \begin{array}{ccc}
c_l & s_l & 0 \\
- s_l & c_l & 0 \\
0 & 0 & 1 
\end{array} \right)
\left( \begin{array}{ccc}
e^{i \varphi} & 0 & 0 \\
0 & c & s \\
0 & -s & c
\end{array} \right)
\left( \begin{array}{ccc}
c_{\nu} & -s_{\nu} & 0 \\
s_{\nu} & c_{\nu} & 0 \\
0 & 0 & 1
\end{array} \right)
\end{equation}
\noindent
where $s_{\nu}$ stands for $sin \Theta_{\nu}$, $s_l$ for $sin \Theta_l$,
$s$ for $sin \Theta$, etc. The angle $\Theta_{\nu}$ is the solar mixing angle
$\left( \Theta_{sun} \right)$, the angle $\Theta$ the atmospheric angle
$\left( \Theta_{at} \right)$ and the angle $\Theta_l$ describes the mixing
between the electron neutrino and the third mass eigenstate.

The experiments give:
\begin{equation}
\Theta_{\nu} \approx 34^{\circ} \qquad \Theta \approx 65^{\circ} \qquad
\Theta_l < 13^{\circ}
\end{equation}

We assume that the lepton mass matrices are described, like the quarks,
by a texture zero matrix (2). Thus we have:
\begin{eqnarray}
tan \Theta_l & = & \sqrt{\frac{m_c}{m_{\mu}}} \qquad \Theta_l \approx 4^{\circ} 
\nonumber \\
tan \Theta_{\nu} & = & \sqrt{\frac{m_1}{m_2}} \, .
\end{eqnarray}
($m_1, m_2$ are the masses of the first, second neutrino).
Using the experimental value $\Theta_{\nu} \approx 34^{\circ}$, we find:
\begin{equation}
\frac{m_1}{m_2} = \left( tan 34^{\circ} \right)^2 \approx 0.45
\end{equation}
Taking into account the observed (mass)$^2$ differences
$\Delta m_{21}^2 \approx 8 \cdot 10^{-5} eV^2$ and $\Delta m_{32}^2 \approx
2.3 \cdot 10^{-3} eV^2$, we obtain the following neutrino masses:
\begin{eqnarray}
m_1 & \approx & 0.0046 \, \, \, {\rm eV} \nonumber \\
m_2 & \approx & 0.01 \, \, \, {\rm eV}  \nonumber \\
m_3 & \approx & 0.05 \, \, \, {\rm eV}
\end{eqnarray}
 Note that the mass ratio $m_2 / m_3$ is:
\begin{equation}
m_2 / m_3 \cong 0.20
\end{equation}

In order to calculate the atmospheric angle $\Theta$, we take $\phi = \pi$ in
eq. (2) and find:
\begin{eqnarray}
\Theta &\cong & arc tan \sqrt{\frac{m_2}{m_3}} + arc tan
\sqrt{\frac{m_{\mu}}{m_{\tau}}} \nonumber \\
tan \Theta & \cong & \left( \sqrt{\frac{m_2}{m_3}} +
\sqrt{\frac{m_{\mu}}{m_{\tau}}} \right) / \left( 1 -
\sqrt\frac{m_2}{m_3} \cdot \frac{m_{\mu}}{m_{\tau}} \right)
\, .
\end{eqnarray}
The angle $\Theta$ is about 38$^{\circ}$. In eq. (12) there is a phase
parameter $e^{i \phi}$ multiplying the second term. In order to obtain
38$^{\circ}$, we have to assume that $\phi$ is close to zero, i. e. in the
leptonic sector the CP violation should be very xmall, at least one order
of magnitude smaller than for the quarks.

Our expected value $\Theta \approx 38^{\circ}$ is on the low side of the
experimental data, which give $\Theta \approx 45^{\circ} \pm 7^{\circ}$.
We have $sin^2 \Theta \approx 0.94$.

The matrix element $V_{3e}$ of the mixing matrix is $sin \Theta_l \cdot
sin \Theta$, and we find: $V_{3e} \approx 0.043$. New experiments with
reactor neutrinos might detect this matrix element.

We summarize: We assume that the lepton mass matrices have the texture
zero form (3) with $D = 0$. The three mixing angles can be calculated
as functions of the lepton masses.

The three neutrino masses are
\begin{eqnarray}
m_1 & \approx & 0,005 \, \, \, {\rm eV}, m_2 \approx 0,01 \, \, \, {\rm eV},
\nonumber \\ m_2 & \approx & 0,05 \, \, \, {\rm eV}.
\end{eqnarray}
There is a normal mass hierarchy: $m_1 < m_2 < m_3$, but the mass
ratios 0.5 and 0.7 are much smaller than the mass ratios for the
quarks ($u$--quark: 0.005, 0.006; $d$--quarks: 0.05, 0.04).

We obtain:
\begin{eqnarray}
tan \Theta_{\nu} & = & tan \Theta_{sun} = \sqrt{m_1 / m_2} \nonumber \\
tan \Theta_l & = & \sqrt{m_e / m_{\mu}} \nonumber \\
\Theta & = & \Theta_1 + \Theta_2 \nonumber \\
tan \Theta_1 & = & \sqrt{m_2 / m_3 } \nonumber \\
tan \Theta_2 & = & \sqrt{m_{\mu} / m_{\tau}} \, .
\end{eqnarray}

We do expect that the CP violation
in the lepton sector is much smaller than the CP--violation in the quark
sector. Thus in the coming experiments it will not be possible to observe a
CP--violation.


\end{document}